\documentclass[preprint,12pt,3p,number,sort&compress]{elsarticle}

\usepackage{graphicx}

\usepackage{amssymb}
\usepackage{amsthm}
\usepackage{mathrsfs}
\usepackage[fleqn]{amsmath}
\usepackage{hyperref}
\usepackage{braket}
\setcitestyle{square}
\usepackage{color}

\biboptions{comma,round}

\biboptions{}

\begin{document}

\begin{frontmatter}

\title{Qubit assisted enhancement of quantum correlations in an optomechanical system}

\author[label1]{Subhadeep Chakraborty\fnref{label3}}
\address[label1]{ Department of Physics, Indian Institute of Technology  Guwahati, Guwahati-781039, Assam, India}

\cortext[cor1]{I am Corresponding author}

\ead{c.subhadeep@iitg.ernet.in}

\author[label1]{Amarendra K. Sarma\corref{cor1}}
\ead{aksarma@iitg.ernet.in}


\begin{abstract}
We perform a theoretical study on quantum correlations in an optomechanical system where the mechanical mirror is perturbatively coupled to an auxiliary qubit. In our study, we consider logarithmic negativity to quantify the degree of stationary entanglement between the cavity field and mechanical mirror, and, Gaussian quantum discord as an witness of the quantumness of the correlation beyond entanglement. Utilizing experimentally feasible parameters, we show that both entanglement and quantum discord enhance significantly with increase in mirror-qubit coupling. Moreover, we find that in presence of the mirror-qubit coupling entanglement could be generated at a considerably lower optomechanical coupling strength, which is also extremely robust against the environmental temperature. Overall, our proposed scheme offers some considerable advantages for realizing the continuous-variable quantum information and communication.
\end{abstract}


\end{frontmatter}


\section{Introduction}
\label{sec1}
Observing the fundamental quantum behavior of a macroscopic object has been a long sought-after goal in quantum physics. This is because, it not only leads to a deeper understanding to the quantum states residing at the border between classical and quantum world but also allows to apply cutting-edge quantum technologies to macroscale devices. In this regard, owing to the recent advances of microfabrication and nanotechnology, cavity optomechanical system has preponderantly emerged as a versatile platform to realize the quantum regime of a macroscopic mechanical oscillator. Substantial progress has been made in this context, both in theory and experiment. Current researches in this area mainly include cavity cooling of a mechanical oscillator \cite{Schliesser,wilson,Marquardt_cool,genes_cool,teufel_cool,chan_cool,bijita}, continuous-variable entangled state preparation \cite{vitali,Paternostro,genes_ent1,genes_ent2,rogers,barz1,barz2,tian_e,li,nie,ent_ext,my}, quantum squeezing of a mechanical mode \cite{mancini,ian,jahne,Clerk,tan,dall,barz3}, nonclassical state generation \cite{Brooks,Ren,Rabl,rips,akram} and quantum state transfer \cite{zhang,sing,tian,tian2,palomaki} etc. Along with the above mentioned studies, it has also become a key tool for a wide range of potential applications, such as: quantum information processing \cite{man2,rips2,stannigel}, ultra-high precision measurements \cite{regal,krause,teufel_pre,barz4}, gravitational wave detection \cite{grav1,grav2}, emission spectra of a hybrid atom-optomechanical system \cite{mirza} and even for biological measurements \cite{taylor}. Overall, preparation and manipulation of the quantum states of a macroscopic mechanical oscillator are now intensively pursued for different fundamental and technological aspects of cavity optomechanics.

A typical cavity optomechanical system \cite{rev_aspel} consists of a laser driven optical cavity with a movable end mirror. So far, in most of the studies, the mechanical mirror is treated as a pure quantum harmonic oscillator. However, regarding the quantumness of the mechanical mirror, it should be noted that a pure quantum harmonic motion is dynamically analogous to its classical counterpart, as, the expectation values of the canonical observables obey the same set of classical equation of motion \cite{ehrenfest}. Therefore, to fully explore the quantum nature of the mechanical mirror, it may be useful to introduce additional nonlinearity in the system. Here, it is worth mentioning that the intrinsic nonlinearity of a sub-gigahertz mechanical oscillator is usually very small (nonlinear amplitude smaller than $10^{-15}\omega_m$ \cite{intr_nonlinear}), and, therefore is relevant only in the regime of large oscillation amplitudes. Several studies have been reported to generate nonlinearities that are strong in the quantum regime \cite{Xiang,tian_def,kurt,armour}. In particular, recently, Rips et al. \cite{rips3} have proposed a scheme to enhance the intrinsic geometrical nonlinearity via an inhomogeneous electrostatic field. This has been further exploited by Djorw\'{e} et al.  in Ref. \cite{djorwe} to generate robust optomechanical entanglement. Along this line, in 2009 Jacobs et al. \cite{jacobs} have reported a theoretical study to engineer wide range of nonlinearities in a mesoscopic  resonator by coupling it to a simple auxiliary system. Following \cite{jacobs}, L\"{u} et al. \cite{lu} engineered a Duffing nonlinearity to study mechanical squeezing in optomechanical system, and, Wang et al. \cite{wang} constructed a cubic nonlinearity to investigate steady state mechanical squeezing in a hybrid atom-optomechanical system. Furthermore,  in Ref. \cite{jacobs2} Jacobs. et al. showed that it is possible to engineer superposition states by coupling a qubit to the square of the resonator's position. A plethora of experimental studies on the coupling between the mechanics and a two-level system can also be found Refs. \cite{mq1,mq2,mq3,mq4,mq5,mq6}.

Motivated by these works, we consider a hybrid optomechanical system in which the mechanical mirror is perturbatively coupled to a single auxiliary qubit. Following \cite{jacobs}, we find that this coupling results in an additional second-order term $q^2$, in the Hamiltonian describing the system where $q$ is the position of the mechanical mirror. All the higher order nonlinear terms are neglected, as we assume weak coupling between the mechanical mirror and the qubit. First, within the standard linearized description we study the stationary entanglement between the cavity field and mechanical mirror. Then, to investigate beyond entanglement we extend our calculation to quantum mutual information and Gaussian quantum discord \cite{d1,d2,d3,d4,d5,d6}. The remaining of the paper is organized as follows: In Sec. \ref{sec2}, we introduce the optomechanical system and derive the dynamics of the quantum fluctuations. In Sec. \ref{sec3}, we focus on the steady-state optomechanical entanglement and study its dependence on the mirror-qubit coupling. Finally, in Sec. \ref{sec4}, we discuss the effect of the mirror-qubit coupling on the quantum mutual information and Gaussian quantum discord, followed by a conclusions in Sec. \ref{sec5}.
\begin {figure}[t]\label{1}
\begin {center}
\includegraphics [height=3.75cm, width =8.0 cm]{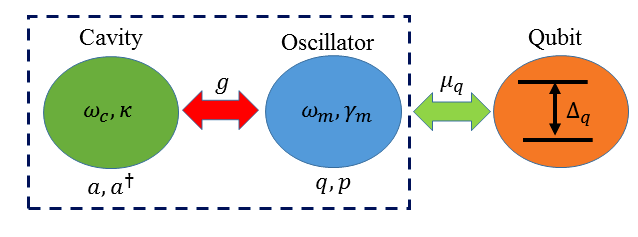}
\caption {(Color online) Schematic diagram of the cavity optomechanical system which consists of an optical cavity with frequency $\omega_c$ coupled to to a mechanical mirror of frequency $\omega_m$, via radiation pressure. Furthermore, we consider the mechanical mirror to be perturbatively coupled to an auxiliary qubit.}
\end{center}
\end{figure}

\section{System Dynamics}\label{sec2}

As depicted in Fig. 1, we consider a hybrid optomechanical system where the mirror is perturbatively coupled to an auxiliary qubit. The Hamiltonian of this system can be written as (in the unit of $\hbar=1$):
\begin{gather}
H=H_0+H_{mq}, \label{Hamiltonian} \\
H_0=\omega_c a^\dagger a+\frac{\omega_m}{2}(p^2+q^2)-ga^\dagger a q,\\
H_{mq}=-\frac{\eta}{2}q^2,
\end{gather}
where $a$ is the annihilation operator of the cavity field (with frequency $\omega_c$), $q$ is the dimensionless position operator of the mechanical mirror (with frequency $\omega_m$), and $g$ is the single-photon optomechanical coupling rate. In Eq. \ref{Hamiltonian}, $H_0$ describes the Hamiltonian of the conventional optomechanical system, while $H_{mq}$ refers to the Hamiltonian for the mirror-qubit interaction with $\eta$ being the coupling strength (see Appendix for a brief discussion on the generation of this term).

In addition, the system dynamics includes the fluctuation-dissipation processes affecting both the cavity field and mechanical mirror. Starting from Hamiltonian \ref{Hamiltonian}, the time evolution of the system is given by the following set of nonlinear quantum Langevin equations:
\begin{subequations}
\begin{gather}
\dot{q}=\omega_m p,\\
\dot{p}=-\omega_m q+\eta q-\gamma_m p+ga^\dagger a+\xi,\\
\dot{a}=-(\kappa+i\Delta_0)a+igaq-iE_0+\sqrt{2\kappa_{ex}}a_{in}+\sqrt{2\kappa_0}f_{in},\label{int_loss}
\end{gather}
\end{subequations}
where $\Delta_0=\omega_c-\omega_l$ denotes the detuning of the cavity field from the laser frequency $\omega_l$, $\kappa=\kappa_{ex}+\kappa_0$ is the total decay rate of the cavity \cite{rev_aspel} and $\gamma_m$ is the mechanical damping rate.We note that in Eq. \ref{int_loss}, while defining the total cavity decay rate $\kappa$, we have considered two distinct contributions, one arising due to the losses at the input cavity mirror $\kappa_{ex}$ and a second contribution from the internal losses of the cavity $\kappa_0$, associated with the scattering and absorption losses behind the first mirror. The input noises acting on the cavity field and mechanical mirror are characterized by the following set of non-zero correlation functions \cite{qop,brown}, given by
\begin{subequations}
 \begin{gather}
 \langle a_{in}(t)a^\dagger _{in}(t^\prime)\rangle=\delta(t-t^\prime),\\
 \langle f_{in}(t)f^\dagger _{in}(t^\prime)\rangle=\delta(t-t^\prime),\\
 \langle\xi(t)\xi(t^\prime)\rangle=\frac{\gamma_m}{\omega_m}\int\frac{d\omega}{2\pi}e^{-i\omega(t-t^\prime)}\omega\left[\coth\left(\frac{\hbar \omega}{2K_BT}\right)+1\right],
 \end{gather}
\end{subequations}
where $K_B$ is the Boltzmann constant and $T$ is the mirror temperature. However, in the limit of large mechanical quality factor $Q=\omega_m/\gamma_m\gg1$, one could well approximate the Brownian noise $\xi(t)$ to a Markovian delta-correlated relation \cite{delta_markov}:
\begin{align}
\langle\xi(t)\xi(t^\prime)+\xi(t^\prime)\xi(t)\rangle /2\simeq\gamma_m\left(2n_{th}+1\right)\delta\left(t-t^\prime\right),
\end{align}
with $n_{th}=\left[\mathrm{ exp}\left(\frac{\hbar \omega_m}{K_BT}\right)-1\right]^{-1}$ being the mean thermal phonon number.

For an intensely driven cavity, we can further expand each Heisenberg operators as a sum of its $c$-number classical steady-state value plus an additional small fluctuation operator with zero mean value: $a=a_s+\delta a$, $q=q_s+\delta q$, $p=p_s+\delta p$. The steady-state values are obtained by solving the following nonlinear algebraic equations:
\begin{subequations}
\begin{gather}
p_s=0,\\
\omega_mq_s-\eta q_s-g|a_s|^2=0,\\
(\kappa+i\Delta)a_s-\sqrt{2\kappa_{ex}}\alpha_{in}=0,
\end{gather}
\end{subequations}
where $\alpha_{in}=\langle a_{in}\rangle$ and $\Delta=\Delta_0-gq_s$ is the effective cavity detuning, modified owing to radiation pressure interaction. On the other hand, the dynamics of the quantum fluctuations are linearized in the limit $|a_s|\gg1$ and given by:
\begin{subequations}\label{fluc}
\begin{gather}
\dot\delta q=\omega_m\delta p,\\
\dot\delta p=-\left(\omega_m-\eta\right)\delta q-\gamma_m \delta p+\frac{G}{\sqrt2}\left(\delta a+\delta a^\dagger\right)+\xi,\\
\dot\delta a=-(\kappa+i\Delta)\delta a+i\frac{G}{\sqrt2}\delta q+\sqrt{2\kappa_{ex}}a_{in}+\sqrt{2\kappa_0}f_{in}.
\end{gather}
\end{subequations}
Here $a_s$ is considered to be real and $G=ga_s\sqrt2$ is the effective optomechanical coupling strength, enhanced by the intracavity field amplitude.

Next, we introduce the dimensionless quadrature operators, respectively, for the cavity field and the corresponding Hermitian input noise operators, as follows:
$\delta X\equiv\frac{\left(\delta a+\delta a^\dagger\right)}{\sqrt2}$, $\delta Y\equiv\frac{\left(\delta a-\delta a^\dagger\right)}{i\sqrt2}$, and
$X_{in}\equiv\frac{\left(a_{in}+a_{in}^\dagger\right)}{\sqrt2}$, $Y_{in}\equiv\frac{\left(a_{in}-a_{in}^\dagger\right)}{i\sqrt2}$, $F_{in}\equiv\frac{\left(f_{in}+f_{in}^\dagger\right)}{\sqrt2}$, $G_{in}\equiv\frac{\left(f_{in}-f_{in}^\dagger\right)}{i\sqrt2}$. With the above definitions, we express Eq. \ref{fluc} in a matrix form:
\begin{align}
\dot u(t)=Au(t)+n(t), \label{dot_u}
\end{align}
where $u^T(t)=\left(\delta q(t),\delta p(t),\delta X(t),\delta Y(t)\right)$ is the vector of continuous-variable (CV) fluctuation operators, $A$ is the drift matrix:
\begin{align}
A=\left(
  \begin{array}{cccc}
    0 & \omega_m & 0 & 0 \\
    -\left(\omega_m-\eta\right) & -\gamma_m & G & 0 \\
    0 & 0 & -\kappa & \Delta \\
    G & 0 & -\Delta & -\kappa \\
  \end{array}
\right),
\end{align}
and $n^T(t)=\left(0,\xi(t),\sqrt{2\kappa_{ex}}X_{in}(t)+\sqrt{2\kappa_0}F_{in}(t),\sqrt{2\kappa_{ex}}Y_{in}(t)+\sqrt{2\kappa_0}G_{in}(t)\right)$ is the vector of corresponding noises. The solution of Eq. \ref{dot_u} reaches its steady state when all the eigenvalues of the drift matrix $A$ have negative real parts. These stability conditions are checked via the Routh-Hurwitz criteria \cite{RH}, which are given in terms of the system parameters, by the following two nontrivial equations:
\begin{subequations}
\begin{multline}
2\gamma_m[\Delta^4+\Delta^2(\gamma_m^2+2\gamma_m\kappa+2\kappa^2-2\omega_m^2)+(\gamma_m\kappa+\kappa^2+\omega_m^2)^2]\\+G^2\Delta\omega_m(\gamma_m+2\kappa)^2+4\eta\gamma_m\kappa\omega_m(\Delta^2-\gamma_m\kappa-\kappa^2-\omega_m^2)+2\eta^2\omega_m^2\gamma_m\kappa>0,
\end{multline}
\begin{align}
\left(\Delta^2+\kappa^2\right)\omega_m^2-\eta\left(\Delta^2+\kappa^2\right)\omega_m-G^2\Delta\omega_m>0. \label{thres}
\end{align}
\end{subequations}
\begin{figure}[!b]\label{2}
\centering
\includegraphics[height=4.55cm, width =12.5cm]{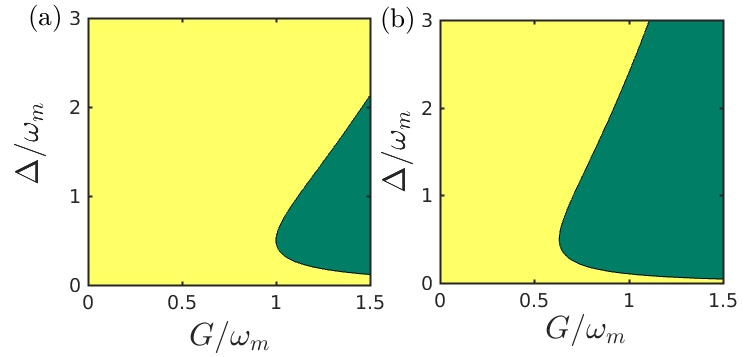}
\begin{flushleft}
\caption {(Color online) Stable (yellow) and unstable (green) region of the optomechanical system in the $(G/\omega_m,\Delta/\omega_m)$ plane, respectively for (a) $\eta/\omega_m=0$ and (b) $\eta/\omega_m=0.6$. The other parameters are chosen to be $\omega_m/(2\pi)=10$ MHz, $Q=10^5$, $\kappa/(2\pi)=5$ MHz and $T_0=0.6$ K.}
\end{flushleft}
\end{figure}

It should be noted that, strictly for red-detuned driving the first condition is always satisfied, only the second condition becomes relevant. Following a simplification of Eq. \ref{thres}, we obtain
\begin{align}
G<G_{thres}=\sqrt{(\Delta^2+\kappa^2)(\omega_m-\eta)/\Delta}.
\end{align}
where $G_{thres}$ refers to the threshold optomechanical coupling strength, separating the stable and unstable phase of the system. We note that in contrast to standard optomechanical system \cite{vitali}, now the system's stability depends on the strength of the mirror-qubit coupling and one must have $\eta<\omega_m$. To numerically illustrate our result, we consider a set of experimentally accessible parameters as implemented in Refs. \cite{gigan,Arcizet}. Figs. 2(a) and (b) depict the stable (yellow) and the unstable (green) phase of the optomechanical system in the $(G/\omega_m,\Delta/\omega_m)$ plane, respectively without ($\eta/\omega_m=0$) and with the mirror-qubit coupling  ($\eta/\omega_m=0.6$). It is clear that in the former case the stability range is much wider than the later, i.e. the system remains stable from $G/\omega_m=0.0$ to $G/\omega_m\approx1.0$ (Fig. 2(a)). On the contrary, in the presence of the mirror-qubit coupling $\eta/\omega_m=0.6$ (Fig. 2(b)), the stability region stretches from $G/\omega_m=0.0$ to $G/\omega_m\approx0.63$. Therefore, we can infer that the mirror-qubit coupling imposes a serious limitation on the maximum allowable effective optomechanical coupling strength. In the remainder of the paper, to operate safely within the stability range of the system, we will restrict ourselves to an optomechanical coupling strength $G/\omega_m\leq0.6$.

\section{Optomechanical Entanglement}\label{sec3}
Due to the above linearized dynamics and the zero-mean Gaussian nature of the quantum noises $a_{in}$ ($f_{in}$) and $\xi$, the steady state for the quantum fluctuations is a zero-mean Gaussian bipartite state, which is fully characterized by its $4\times4$ correlation matrix (CM):
\begin{align}
V_{ij}=\left(\langle u_i(\infty)u_j(\infty)+u_j(\infty)u_i(\infty)\rangle\right)/2.
\end{align}
Here, $u^T(\infty)=\left(\delta q(\infty),\delta p(\infty),\delta X(\infty),\delta Y(\infty)\right)$ is the vector of CV fluctuation operators in the steady-state $(t\rightarrow\infty)$.
When the system is stable, the steady-state CM satisfies the following Lyapunov equation \cite{vitali}:
\begin{align}
AV+VA^T=-D,
\end{align}
where $D$ is a diagonal matrix  of the matrix of noise correlations:
\begin{align}
D=\mathrm{Diag}\left[0,\gamma_m(2n_{th}+1),\kappa,\kappa\right].
\end{align}
Now, to study the quantum entanglement and the other related quantities, we express the above CM $V$ in a $2\times2$ block form:
\begin{align}
V=\left(
  \begin{array}{cc}
    V_m & V_{mc} \\
    V_{mc}^T & V_c \\
  \end{array}
\right),
\end{align}
where $V_m$ and $V_c$, respectively, desscribe the local properties of the mechanical mirror and the cavity field, while $V_{mc}$ refers to non-local correlation between them. Moreover, we consider the following set of four symplectic invariants derived from the CM $V$:
\begin{align}
I_1=\mathrm{det}V_m, \quad I_2=\mathrm{det}V_c, \\\nonumber I_3=\mathrm{det}V_{mc}, \quad I_4=\mathrm{det}V.
\end{align}
The degree of quantum entanglement between the mechanical mirror and cavity field can be determined by calculating the so-called logarithmic negativity \cite{vidal,adesso}, defined as:
\begin{align}
E_N=\mathrm{max}\left[0,-\ln2\tilde{\nu}^-\right],
\end{align}
where $\tilde{\nu}^-\equiv2^{-1/2}\left[\tilde{\Delta}(V)-\sqrt{\tilde{\Delta}(V)^2-4I_4}\right]^{1/2}$ is the lowest symplectic eigenvalue of the partial transpose of $V$ with $\tilde{\Delta}(V)\equiv I_1+I_2-2I_3$. Therefore, the Gaussian state is entangled if and only if $E_N>0$, which is equivalent to  $\tilde{\nu}^-<1/2$ (Simon's necessary and sufficient criteria \cite{simon}).
\begin{figure}[t]\label{3}
\centering
\includegraphics[height=5.0cm, width =14cm]{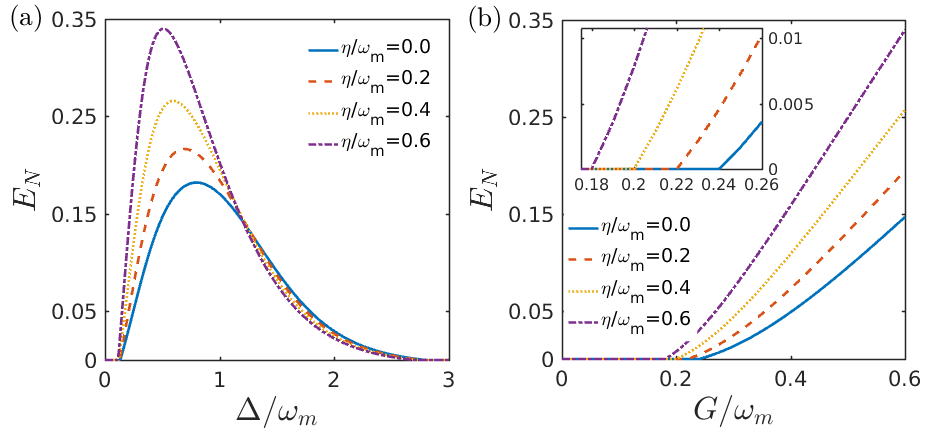}
\caption {(Color online) Plot of logarithmic negativity $E_N$ as a function of (a) normalized detuning $\Delta/\omega_m$ for a fixed coupling strength $G/\omega_m=0.6$ and (b) dimensionless effective optomechanical coupling strength $G/\omega_m$ at a fixed detuning $\Delta/\omega_m=0.5$, with all the other parameters same as in Fig. 2.}
\end{figure}

In Fig. 3(a), we plot logarithmic negativity $E_N$ versus normalized detuning $\Delta/\omega_m$, for different mirror-qubit coupling strengths. It shows that for the chosen parameters, the usual system ($\eta=0$) exhibits stationary optomechanical entanglement, which reaches its maximum $E_n^{max}=0.18$ around $\Delta/\omega_m=0.8$. However, the introduction of the mirror-qubit coupling enhances the degree of entanglement and shifts the position of the maxima towards a lower cavity detuning. In particular, we find a high degree of entanglement $E_n^{max}=0.34$ at $\Delta/\omega_m=0.51$ for $\eta/\omega_m=0.6$. Here, it is also worth mentioning that for the chosen parameters, $(\Delta/\omega_m,\eta/\omega_m=0.80,0.0)$ and $(\Delta/\omega_m,\eta/\omega_m=0.51,0.6)$ respectively yield: $G_{thres}=1.03\omega_m$ and $G_{thres}=0.63\omega_m$. This implies that the maximum entanglement appears only when $G\rightarrow G_{thres}$ i.e. for the maximum possible optomechanical coupling strength. Therefore, we infer that this entanglement enhancement is achieved due to the additional mirror-qubit coupling, which brings the system close to the instability. Fig. 3(b) depicts the same logarithmic negativity $E_N$ as a function of the dimensionless effective optomechanical coupling $G/\omega_m$, for different mirror-qubit coupling strengths. As expected, after a critical coupling strength $(G_c)$, entanglement appears and increases with increasing optomechanical coupling. We note that this enhancement becomes more profound with an increase in mirror-qubit coupling strength $\eta/\omega_m$. More importantly, as illustrated in the inset of Fig. 3(b), we observe a lowering of this critical coupling strength $G_c$, with an increase in $\eta/\omega_m$. For example, one can see that the usual system $(\eta=0)$ starts exhibiting entanglement at $G_c/\omega_m=0.24$, whereas, in presence of the mirror-qubit coupling $(\eta/\omega_m=0.6)$ it shows entanglement at $G_c/\omega_m=0.18$. Hence, the mirror-qubit coupling not only enhances the degree of entanglement but also generates it at a considerably lower optomechanical coupling strength.

Next, to study the robustness of our scheme, we plot in Fig. 4 logarithmic negativity $E_N$ with  respect to the mean thermal phonon number $n_{th}$, for various mirror-qubit coupling strengths. We observe that the degree of entanglement decreases monotonically with increasing thermal phonon number. However, the maximum number of thermal phonons up to which entanglement persists increases with increasing mirror-qubit coupling strength. For example, in the generic case $(\eta=0)$ the entanglement is maintained up to $n_{th}=6626$ ($T=3.18$ K). In contrast, in presence of the mirror-qubit coupling $\eta=0.6\omega_m$ the entanglement persists up to a relatively high $n_{th}=11752$ ($T=5.64$ K). Thus, the bipartite entanglement in presence of the mirror-qubit coupling becomes more robust against thermal phonon fluctuations.
\begin{figure}[!h]\label{4}
\centering
\includegraphics[height=5cm, width =6cm]{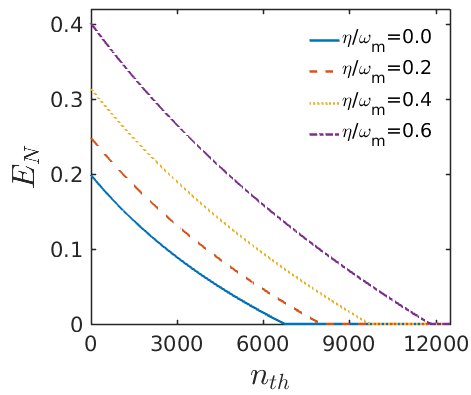}
\begin{flushleft}
\caption {(Color online) Plot of logarithmic negativity $E_N$ as a function of the mean thermal phonon number $n_{th}$. Here, the effective cavity detuning and optomechanical coupling strength are respectively fixed at $\Delta/\omega_m=0.5$ and $G/\omega_m=0.6$, with all the other parameters same as in Fig. 2.}
\end{flushleft}
\end{figure}
\section{Quantum mutual information and Gaussian quantum discord}\label{sec4}
Besides quantum entanglement, quantum discord is a recently proposed measure of the quantum correlation between two systems. In particular, it is now well established that even a separable state can posses some amount of quantum correlation, which is characterized by a nonzero discord. Therefore, it will be quite interesting to study the effect of the mirror-qubit coupling on a more genuine measure of the quantum correlations. By definition, quantum discord is given by the following difference (with respect to a measurement made on the cavity mode $a$):
\begin{align}
\mathscr{D}_G(m|c)=I_M(mc)-C(m|c),
\end{align}
where, the first term $I_M(mc)$ quantifies the total amount of correlation shared by the cavity field and mechanical mirror, namely by quantum mutual information, while  the second term $C(m|c)$ refers to the one-way classical correlation between the cavity field and mechanical mirror. For the considered Gaussian bipartite system characterized by the correlation matrix $V$, the quantum mutual information is written as \cite{Olivares}
\begin{align}
I_M(mc)=f(\sqrt{I_1})+f(\sqrt{I_2})-f(\nu^+)-f(\nu^-),
\end{align}
where, the function $f$ is defined as:
\begin{align}
f(x)\equiv(x+\frac{1}{2})\ln(x+\frac{1}{2})-(x-\frac{1}{2})\ln(x-\frac{1}{2}),
\end{align}
and,
\begin{align}
\nu^\pm\equiv2^{-1/2}\left[\Delta(V)\pm\sqrt{\Delta(V)^2-4I_4}\right]^{1/2},
\end{align}
are the two symplectic eigenvalues of $V$ with $\Delta(V)\equiv I_1+I_2+2I_3$. On the other hand, Gaussian quantum discord is evaluated by performing Gaussian measurement only and expressed as \cite{Olivares}:
\begin{align}
\mathscr{D}_G(m|c)=f(\sqrt{I_2})-f(\nu^+)-f(\nu^-)+f(\sqrt{W})
\end{align}
where, the function $f$ is defined in Eq. (24) and
\begin{align}
W=\begin{cases}
    \left[\frac{2|I_3|+\sqrt{4I_3^2+(4I_2-1)(4I_4-I_1)}}{(4I_2-1)}\right]^2 \quad \mathrm{if} \frac{4(I_1I_2-I_4)^2}{(I_1+4I_4)(1+4I_2)I_3^2}\leq1, \\\;
    \frac{I_1I_2+I_4-I_3^2-\sqrt{(I_1I_2+I_4-I_3^2)^2-4I_1I_2I_4}}{2I_2} \quad \mathrm{otherwise}.
\end{cases}
\end{align}

In Figs. 5(a) and (b), we respectively plot the quantum mutual information $I_M$ and the Gaussian quantum discord $\mathscr{D}_G$ as a function of the normalized detuning $\Delta/\omega_m$, for different mirror-qubit coupling strengths. In both the figures, we find that for the generic system ($\eta/\omega_m=0$) $I_M$ and $\mathscr{D}_G$ are maximum near $\Delta/\omega_m=0$, and, then decrease monotonically with increasing $\Delta/\omega_m$ \cite{Brunelli}. However, with the introduction of mirror qubit coupling we observe a completely different behavior: Depending the strength of mirror-qubit coupling, both $I_M$ and $\mathscr{D}_G$ enhance significantly at higher cavity detuning. In particular, for $\eta/\omega_m=0.6$ we obtain the maximum $I_M=1.29$ at $\Delta/\omega_m=0.49$ and $\mathscr{D}_G=0.20$ at $\Delta/\omega_m=0.80$. Therefore, this along with Fig. 3, confirms that invoking the mirror-qubit coupling to a generic optomechanical system is a promising mean to improve the quantum correlations between the cavity field and mechanical mirror.
\begin{figure}[t]\label{5}
\centering
\includegraphics[height=5.0cm, width =14cm]{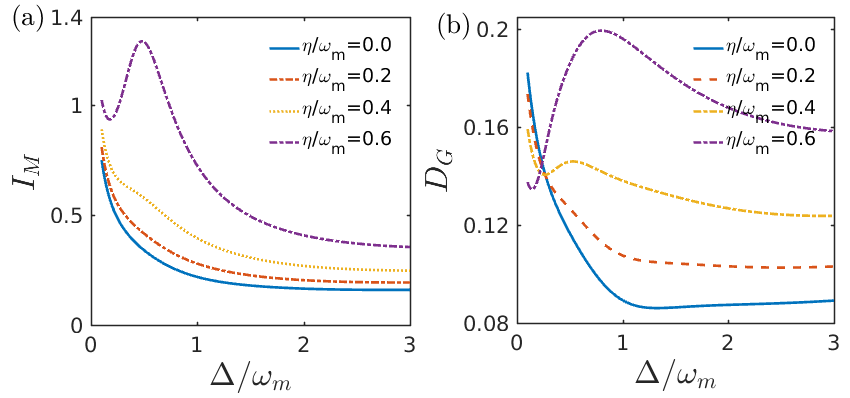}
\caption {(Color online) Plot of (a) quantum mutual information $I_M$ and (b) Gaussian quantum discord $\mathscr{D}_G$ as a function of the normalized detuning $\Delta/\omega_m$, for different mirror-qubit coupling strengths $\eta/\omega_m$, with all the other parameters same as in Fig. 2.}
\end{figure}
\section{Conclusion}\label{sec5}
In conclusion, we reported a scheme to enhance the quantum correlation between a light field and a mechanical mirror. The system under consideration, consists of an optomechanical system where the mechanical mirror is perturbatively coupled to an auxiliary qubit. Here, it is worth mentioning that in order to maintain the perturbation approximation valid, the maximum eigenvalue of the mirror-qubit interaction should be significantly smaller than the separation energy of the auxiliary qubit. Therefore, while driving the cavity, one should consider a suitable driving amplitude, so that the qubit is not pushed into a nonlinear regime. Within the standard linearized description, we first showed that due to the mirror-qubit coupling, the system becomes unstable at a considerably lower optomechanical coupling strength leading to significant enhancement of steady-state entanglement near the instability threshold. Moreover, we found that in presence of the mirror-qubit coupling, the entanglement could be generated at lower optomechanical coupling strength which is extremely robust against the mirror temperature. Then, to investigate the correlations which can not be captured by studying entanglement only, we extended our calculation to quantum mutual information and Gaussian quantum discord. We observed that unlike the generic system (without the mirror-qubit coupling), now both the mutual information and the quantum discord enhance significantly at a higher cavity detuning with an increase in the mirror-qubit coupling strength. As all the chosen parameters are achievable within the current state-of-the-art experimental setups, our proposed scheme could be implemented to improve the quantum correlations in optomechanical platforms.

\section*{Acknowledgement}
S. Chakraborty would like to acknowledge the financial support from Ministry of Human Resource development (MHRD), Government of India.

\section*{Appendix}\label{app}
Here, we provide a brief discussion on the generation of the additional second-order term, introduced in the system Hamiltonian. As prescribed in Ref. \cite{jacobs}, we couple the mechanical mirror to a single auxiliary qubit with the Hamiltonian $H_{q}=(\Delta_{q}/2)\sigma_x$, via a linear interaction $H_{int}=\mu_q q\sigma_z$. Here, $\mu_q$ is the interaction strength between the mechanical mirror and auxiliary qubit, and, $q$ be the dimensionless position operator of the mechanical mirror. Then, under the condition $\Delta_q\gg \mu_q$, we use the standard time-independent perturbation theory to diagonalize the auxiliary Hamiltonian: $H_{aux}=H_q+H_{int}$, which gives the eigenvalues of $H_{aux}$, as a power series in $\mu_q q$. Now, by placing the qubit in the eigenstate of $\sigma_x$, we can engineer an additional second-order term $H_{mq}=-2\Delta_q(\mu_q/\Delta_q)^2q^2$ ($\frac{\eta}{2}=2\Delta_q(\mu_q/\Delta_q)^2$) in the Hamiltonian describing the system. Note that in the perturbation expansion, we have ignored all the higher order nonlinear terms in the limit of $\Delta_q\gg \mu_q$. It is also worth mentioning that as reported in Ref. \cite{jacobs}, the errors induced by qubit dephasing  on the generated perturbation coefficients are very small, and hence could be neglected. 


\bibliographystyle{elsarticle-num}


\end{document}